# YOUNG STELLAR GROUPS IN M 33 GALAXY: DELINEATION AND MAIN PARAMETERS

LUBA VASSILEVA[1], TODOR VELTCHEV[2], TSVETAN GEORGIEV[1]
and PETKO NEDIALKOV[2]

[1]*Institute of Astronomy, Bulgarian Academy of Sciences,*
*72 Tsarigradsko Chaussee Blvd., 1784 Sofia, Bulgaria*
e-mail: tsgeorg@libra.astro.bas.bg
[2]*Department of Astronomy, Faculty of Physics, Sofia University,*
*5 James Bourchier Blvd., 1164 Sofia, Bulgaria*
e-mail: eirene@phys.uni-sofia.bg

**Abstract.** The problem of (non-)existence of a typical size of the stellar associations is revisited by use of deep UBVRI stellar CCD photometry in M 33 from the Local Group Survey (Massey et al. 2006). We compare the outlines of the `classical OB associations' (Ivanov 1991) with stellar groups that were selected through an objective method for determination of the local stellar density and delineation. Main parameters of some stellar groups like size, shape and density concentrations are determined.

## 1. INTRODUCTION

The Triangulum spiral galaxy (Messier 33) is a favorite object for study of OB associations because of its inclination angle i=54° (Corbelli and Salucci 2000), relatively large angular size and close distance to us (DM=24.64 ± 0.15, Galleti 2004). The cornerstone was laid by Humphreys and Sandage (1980) who identified 143 associations with typical length scale of 250 pc. Further 289 associations were outlined by Ivanov (1987, 1991) on UBV plates taken with the 2m RCC telescop at NAO Rozhen. Some of them turned out to be subgroups of the `associations' in the work of Humphreys and Sandage (1980). The mean size of these young stellar groups is ≈80 pc. Apparently, 47 stellar complexes in M 33 of mean size ≈ 570 pc encompass groups of associations, H II regions and extensive H I clouds (~1.2 kpc) (Ivanov 1991). A thorough investigation of the circumnuclear region of the galaxy (7′ × 10′) was performed by Wilson (1991) using UBV CCD photometry with the 3.6m CFHT and the 60-in Palomar telescope. She found 41 OB associations with mean radii of 40 pc; each of them contains at least 10 blue stars.

Star formation in galaxies occurs on multiple spatial scales: from stellar complexes (sizes of >100 pc) through OB associations (sizes between 15 and 100 pc) down to individual stellar clusters (sizes < 15 pc) and some substructures within them. Bastian et al. (2007) study stellar structures in M 33 applying an objective method independent on spatial sizes and with effective resolution limit of ~20 pc. Unlike the aforementioned authors, they don't find any typical length scale of the OB associations and conclude that the appearance of such in previous studies was due to a low resolution limit and selection effects. They demonstrate that size distribution of OB association is lognormal and correlates with the spatial distribution of H II regions.

The aim of the present work is to revisit the issue of the typical sizes and shapes of OB associations using a combination of an approach to extract stellar groups from a 2D density field (Sect. 3.1) and a method to delineate them with fitting ellipses (Sect. 3.2).

## 2. STELLAR PHOTOMETRY AND THE SELECTED FIELD

Like Bastian et al. (2007), we use the deep automated UBVRI photometry of M 33 from Local Group Survey (LGS, Massey et al. 2006). The LGS imaging of 3 fields of the galaxy was done with Mosaic CCD camera on KPNO and CTIO 4m- telescopes. It covers about 0.8 $\square^{\circ}$ including all currently active massive star formation regions (Fig. 1a). Although the image quality is modest (median seeing of 1.0"), LGS is a significant contribution in resolving blending effect in dense stellar fields. The sample consists of 28 378 blue massive stars with various reddening (selection criterion (B-V)<1.). Some ~ 15 000 of them were successfully dereddened through the classical Q-method on a color-color diagram (Vassileva et al. 2006).

We chose a frame of $16 \times 11'$, encompassing the circumnuclear region of the galaxy and containing much OB associations (Fig. 1b). The rectangular coordinates are calculated, adopting positional angle *PA*=23$^{\circ}$ and aligning the X axis along the major galaxy axis. Some of the associations (as outlined by Ivanov 1991) are parts of a larger complex and were designated with the small Latin letters, added to its number.

## 3. THE APPLIED METHODS

### 3.1. Selection of dense stellar groups

The approach for extracting dense stellar groups from a 2D discrete field consists of two basic steps: 1) constructing a local density map; and, 2) selecting groups by imposing some fixed low density limit $n^*_{low}$ and a breaking (minimal) distance $d_{break}$ between two different groups. The local density map is built through a rectangular grid. Stellar density was calculated for each grid point within a search radius equal to the grid box size $d_{grid}$. Two neighbour grid points with local stellar density $n^* > n^*_{low}$ are treated to belong to one and the same group while such on the diagonal of a grid box – not (i.e. $d_{break} = d_{grid} \sqrt{2}$). The coordinates of a group center are determined as weighted means of the coordinates of group member points, where the weights of all members are equal in first aproximation.

This Eulerian approach was applied to the whole sample varying 2 free parameters: the search radius $d_{grid} = 3 \div 25)$ and the minimum number of stars within $d_{grid}$: $n^*_{min} = 3 \div 10$, i.e. the minimal local stellar density. Fig. 2 illustrates how this variation influences the number of detached groups. The maximum of the distribution is well pronounced and reflects the typical mean density of the associations.

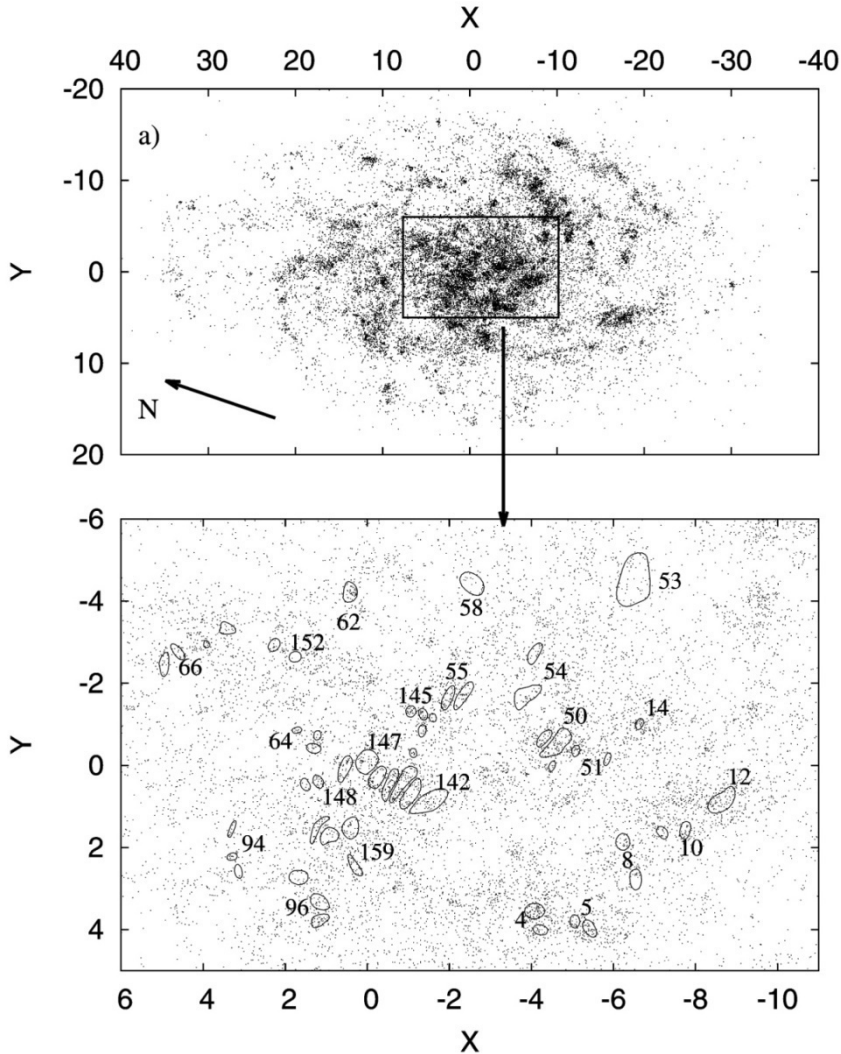

**Figure 1:** The location of the selected blue stars: a) on the plane of M 33; b) within the chosen frame. The outlines of the 'classical OB associations' and their numbers according to Ivanov (1991) are shown. The designations of the subgroups (when present) are not given.

As one could expect, it shifts up with decreasing the minimal local density - as more and more local density fluctuations are selected as 'groups', - and to the left with decreasing $d_{grid}$, i.e. the spatial resolution of the method. We note also a faint secondary maximum to the right that may be interpreted as pointing to a typical stellar density level of the galactic disk. It is clear that below some characteristic $d_{grid}$, determined mainly by the photometric limits, increasingly more groups are omitted. On the other side, real stellar groups merge into larger domains when $d_{grid}$ is about the mean distance between the associations.

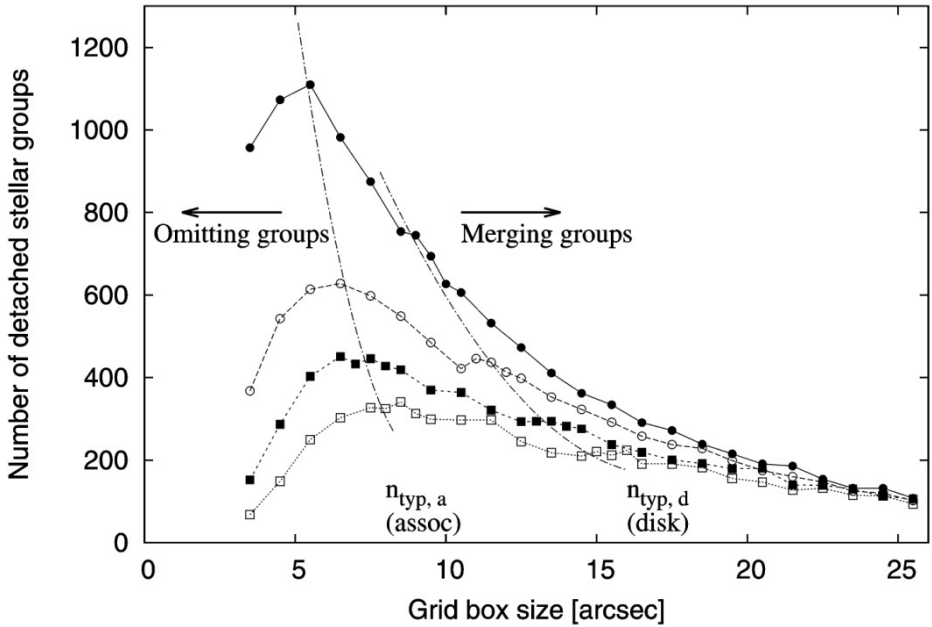

**Figure 2:** Number of selected dense groups as a function of grid box size $d_{grid}$ and the chosen minimal local stellar density: $n_{min}^{*}$=3 (filled circles), $n_{min}^{*}$=4 (open circles), $n_{min}^{*}$=5 (filled squares) and $n_{min}^{*}$=6 (open squares). See text for more details.

### 3.2. Delineation and parameters of stellar groups

The apparent view of a stellar association is a limited 2D distribution on the XY-plane, containing N points ($X_i$, $Y_i$) (i.e. stars). In the general case, it exhibits a pronounced central concentration that could be approximated with a 2D Gaussian. Therefore an elliptical presentation of the apparent shape seems to be useful first approximation. Additional parameter could be introduced to characterize the concentration of the stellar density distribution. The elliptical model includes 5 geometric parameters: center ($X_c$, $Y_c$), semi-axes ($a$, $b$) and the positional angle ($p.a.$) between the major axis and the major galaxy axis (i.e. the OX axis). Meaning and interpretation of the size parameters $a$ and $b$ deserves special attention (see below).

The elliptical parameters could be computed from the 2D moments of the observed discrete distribution. This *method of the moments* was elaborated for the purposes of digital image processing (Sobie 1980), but it is entirely applicable to a discrete 2D distribution.

Another approach, producing the same formulae, consists in finding the ellipse' major axis as result of the orthogonal regression $Y|X$ of the observed distribution. The regression passes through the center of the distribution and minimizes the sum of the squares of the deviations $H_i^2$ measured vertically toward the fitting line $Y(X)=C_X.X+D$:

$$H_i = \frac{|Y_i - C_X.X - D|}{(C_X^2 + 1)^{1/2}}$$

The orthogonal standard deviation is a natural measure of the scatter of the distribution. It may be used as minor semi-axis $b$ of the ellipse while the standard deviation in respect to the perpendicular line to the orthogonal regression $Y|X$, passing through the distribution center, may be used as a major semi-axis $a$. Thus, if one considers a Gaussian 2D distribution, the ellipse semi-axes $a$ and $b$ are estimations of the size parameters $\sigma_1$ and $\sigma_2$ of the Gaussian and the respective ellipse must contain 67% of the Gaussian distribution. From this point of view the method of the moments gives a "$2\sigma$ ellipse'" that must encompass about 96% of the Gaussian distribution. If the 2D distribution is uniform within an ellipse, the method of the moments gives exactly the semi-axes $a$ and $b$ of the ellipse. The so called *concentration index* is defined:

$$CI = \frac{n}{N} \ ,$$

where $n$ is the number of the stars within the $1\sigma$ ellipse and $N$ - the total number of group members. Some characteristic values of this index are: $CI \sim 0.67$ (Gaussian-like distribution), $CI \sim 0.25$ (flat distribution), $CI < 0.25$ (central depression; $CI=0$ means a hole in the center).

## 4. RESULTS

The appearance of a typical mean stellar density for the OB associations $n_{typ,a} \approx 0{,}0020$ pc$^{-2}$, corresponding to the used sample of objects with LGS photometry (Fig. 2), leads to the natural assumption that the outlines of the 'classical associations' (based on photographic photometry) correspond to some typical stellar density as well. Fixing one of the two free parameters in the procedure, described in Sect. 3.1, and varying the other one is equivalent to a variation of $n_{typ,a}$. As seen from Fig. 3a, the found typical mean density reveals different structure of the star formation regions which confirms the conclusions of Bastian et al. (2007) about the dependence of sizes of young stellar groups on photometric limits. For more than twice higher density we find 'cores' that emerge within the outlines of the OB associations.

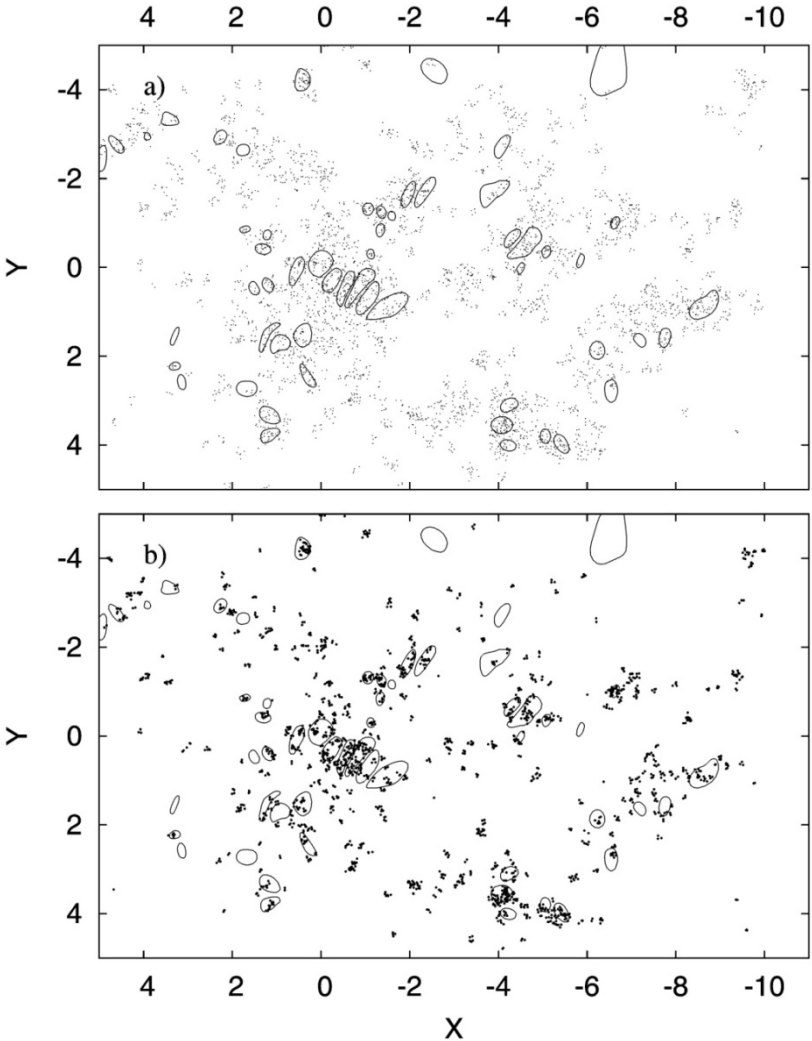

**Figure 3:** Stellar density and the outlines of the 'classical associations': a) for $n_{typ,a}$=0,0020 pc$^{-2}$, corresponding to the maxima in Fig. 2; b) $n_{typ,a}$=0,0050 pc$^{-2}$, exhibiting 'cores' within the associations.

We study such cores and small groups of them, applying the method, described in Sect. 3.2. The main parameters of the elliptical approximations of 6 large complexes are given in Table 1 while in Fig. 4 the shapes and orientations of their substructures ('associations') are juxtaposed with the 'classical' outlines. In all cases the stellar density distribution is rather flat and the outlines of the 'classical associations' seem to detach its local enhancements. Practically all associations contain one or more dense stellar cores. The age segregation within the complexes is a subject for further study. There is no specific trend in the major axis' orientation of the ellipse.

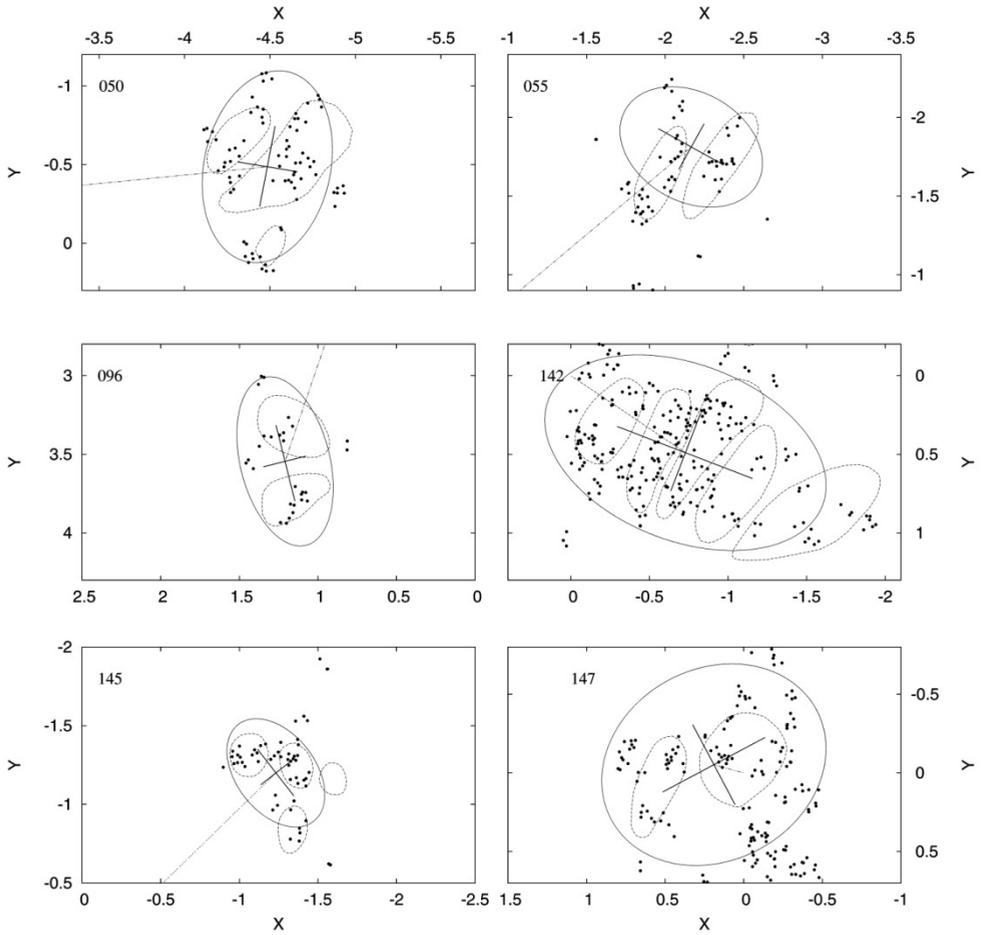

**Figure 4:** Elliptical outlines' approximations (solid) of stellar groups that cover 6 'classical complexes' and their subgroups ('associations'; dashed). The number of each complex according to Humphreys and Sandage (1980) is given. The direction to the center is shown with dashed-dotted line.

**Table 1.** Main parameters of elliptical outline approximations of 6 large complexes. The abbreviations are as follows: $D_c$ - distance to the galaxy center ([arcmin]), $(X_c, Y_c)$ - coordinates of the complex' center, $a$ - major semi-axis ([arcmin]), $d_e = 2\sqrt{a\,b}$ - equivalent diameter ([arcmin]), *p.a.* - positional angle of the major axis, *CI* - concentration index.

| Complex | $D_c$ | $X_c$ | $Y_c$ | $a$ | $b/a$ | $d_e$ | *p.a.* | *CI* |
|---|---|---|---|---|---|---|---|---|
| **050** | 4.511 | -4.484 | -0.486 | 0.616 | 0.603 | 0.956 | 80 | 0.265 |
| **055** | 2.824 | -2.166 | -1.812 | 0.481 | 0.722 | 0.818 | 151 | 0.362 |
| **096** | 3.746 | ~1.209 | ~3.545 | 0.550 | 0.518 | 0.792 | 104 | 0.345 |
| **142** | 0.879 | -0.729 | ~0.490 | 0.934 | 0.601 | 1.449 | 159 | 0.352 |
| **145** | 1.721 | -1.233 | -1.200 | 0.390 | 0.645 | 0.626 | 128 | 0.265 |
| **147** | 0.197 | ~0.190 | -0.052 | 0.740 | 0.824 | 0.902 | 28 | 0.374 |

## Acknowledgement

This research was partially supported by contract Nr. F-201/06 with Scientific Researh Foundation, Ministry of Education and Sciences, Bulgaria.